\documentstyle[aps,graphicx]{revtex}

\begin{document}
\title{{\Large \bf The Woods-Saxon Potential in the Dirac Equation}}
\author{{\large \bf Piers Kennedy}}
\address{Centre for Theoretical Physics, University of Sussex, Brighton 
BN1 9QJ,~UK\\
email: kapv4@pact.cpes.susx.ac.uk}
\maketitle

\begin{quotation}
The two-component approach to the one-dimensional Dirac equation is 
applied to the Woods-Saxon potential. The scattering and bound state 
solutions are derived and the conditions for a transmission resonance (when 
the transmission coefficient is unity) and supercriticality (when the 
particle bound state is at $E=-m$) are then derived. The square 
potential limit is discussed. The recent result that a finite-range symmetric 
potential barrier will have a transmission resonance of zero-momentum 
when the corresponding well supports a half-bound state at $E=-m$ is 
demonstrated.
\end{quotation}

\section*{Introduction}

There is a well-known theorem for low momentum scattering in the
Schr\"{o}dinger equation in one-dimension by an even
potential well  \cite{Senn}, \cite{SdB}: as momentum $k$ tends to zero, 
the reflection
coefficient $L(k)$ tends to unity unless the potential $V(x)$ supports 
a
zero energy resonance \cite{Newton}. In this case $L(k)\rightarrow 0$ 
and correspondingly
the transmission coefficient $T(k)\rightarrow 1$. Bohm \cite{Bohm} 
calls
this a transmission resonance. Recently we \cite{DKC} have generalised 
this
result to the Dirac equation. Since the Dirac equation covers 
anti-particle
scattering as well as particle scattering, the generalisation gives two
distinct results since the $k\rightarrow 0$ limit in the Dirac equation
corresponds both to particle states where the energy $E=m$ and 
anti-particle
states where $E=-m$ and $m$ is the particle mass. The result that
anti-particles can have transmission resonances when they scatter off
potential wells is equivalent to particles having transmission 
resonances
when scattering off potential barriers. This is itself related to the 
result
on barrier penetration found by Klein \cite{Klein} and now called the 
Klein
Paradox.

Our result \cite{DKC} shows that transmission resonances at $k=0$ in 
the
Dirac equation occur for a potential barrier $V=U_{c}(x)$ when the
corresponding potential well $V=-U_{c}(x)$ just supports a bound state 
at $%
E=-m$: this is called a supercritical state. While transmission 
resonances
for a square barrier in the Dirac equation have been known for some 
time 
\cite{CA} and their relationship to supercritical states for a square
well was pointed out more recently \cite{CDI} we are not aware of any 
other
analytic solution of the Dirac equation for which they can be 
demonstrated.
Nevertheless Dosch et al \cite{DJM} showed in 1971 that a transmission
resonance does exist for the Woods-Saxon potential, which is a smoothed 
out
form of the square well/barrier. In this note, we take this example 
further.
We solve the Dirac equation for the Woods-Saxon potential well to
demonstrate the supercritical states and we find complete solutions for 
the
reflection and transmission amplitudes for scattering off a Woods-Saxon
potential barrier. We then are able to demonstrate the correspondence
between the supercritical states and the transmission resonances
analytically. We also consider the limit where the Woods-Saxon 
potential
becomes a square well/barrier.

\section*{The one particle Dirac equation in one dimension}
In the one-dimensional Dirac equation, solutions can be greatly 
simplified by adopting a two-component approach; both the positive and 
negative energy solution states are retained without the added complication of 
spin. Starting with the relativistic free particle Dirac equation 
($\hbar=c=1$):
\begin{equation}
(i \, \gamma^{\mu}\frac{\partial}{\partial x^{\mu}} - m)\psi=0 
\end{equation}
In the presence of an external potential $V(x)$ and taking the gamma 
matrices $\gamma_x$ and $\gamma_0$ to be the Pauli matrices $i \sigma_x$ 
and $\sigma_z$ respectively, the Dirac equation in one dimension can be 
written as
\begin{equation}
(\sigma_x \frac{d}{dx} - (E-V(x)) \sigma_z + m {\bf 1})\psi(x)={\bf 0} 
\end{equation}
The four-spinor, $\psi$, is decomposed into two spinors, $u_1$ and 
$u_2$, so that
\begin{equation}
\psi(x)=\left(
\begin{array}{c}
u_1(x)\\
u_2(x)
\end{array} \right)
\end{equation}
Thus the problem is to solve the coupled differential equations:
\begin{eqnarray}
u_1^{\prime}=-(m+E-V(x))u_2(x)\nonumber \\
u_2^{\prime}=-(m-E+V(x))u_1(x)
\end{eqnarray}
Following a similar procedure to that used by Fl\"ugge \cite{F}, 
introduce the following combinations
\begin{equation}
\phi(x)=u_1(x)+iu_2(x) \qquad , \qquad \chi(x)=u_1(x)-iu_2(x)
\end{equation}
Substituting these into (4) and re-arranging gives:
\begin{eqnarray}
\phi^{\prime}(x)=-im \chi(x) +i(E-V(x))\phi(x)\\
\chi^{\prime}(x)=im \phi(x) -i(E-V(x))\chi(x)
\end{eqnarray}
The two components, $\phi(x)$ and $\chi(x)$, satisfy:
\begin{eqnarray}
\phi^{\prime \prime}(x)+[(E-V(x))^2-m^2+iV^{\prime}(x)]\phi(x)=0\\
\chi^{\prime \prime}(x)+[(E-V(x))^2-m^2-iV^{\prime}(x)]\chi(x)=0
\end{eqnarray}
In the following the full solutions for $\phi(x)$ will be presented. In 
order to establish $\chi(x)$, use will be made of (6).

\section*{The Woods-Saxon Potential}
The Woods-Saxon potential is defined as
\begin{equation}
V(x)= W \left( \frac{\theta (-x)}{1+e^{-a(x+L)}}+\frac{\theta 
(x)}{1+e^{a(x-L)}} \right)
\end{equation}
with $W$ real and positive for a barier or negative for a well; $a$ and 
$L$ are real and positive. $\theta(x)$ is the Heaviside step function.
\begin{figure}[tbph]
\par
\begin{center}
\leavevmode
\includegraphics[width=0.5\linewidth]{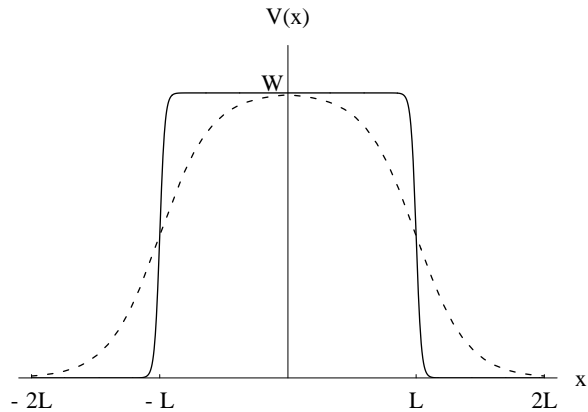} \medskip
\caption{The Woods-Saxon potential barrier for $L=10$ with $a=5$ (solid 
line) and $a=0.5$ (dotted line).}
\end{center}
\end{figure}

At this stage it is worth mentioning that we will be interested in 
potentials where $aL \gg 1$; from the above figure it can be seen that, for 
this condition, the potential has a less pronounced cusp at $x=0$ - the 
potential now closely resembles a square barrier with smooth walls. As 
stated in \cite{DJM} this does not introduce any essential physical 
restriction on the problem and is significant only in that it allows exact 
solutions to be established (albeit in an approximation).

\section*{Scattering States}
First, consider the scattering solutions for $x<0$ with $|E|>m$. On 
making the substitution $y=-e^{-a(x+L)}$, equation (8) becomes
\begin{equation}
a^2 y \frac{d}{dy} \left[ y \frac{d \phi_L}{dy} \right]+ \left[ \left( 
E - \frac{W}{1-y} \right)^2-m^2-\frac{iayW}{(1-y)^2} \right] \phi_L=0
\end{equation}
Splitting off fitting powers of $y$ and$(1-y)$ by setting 
$\phi_L=y^{\mu}(1-y)^{-\lambda}f(y)$ and substituting into the above equation 
reduces it to the hypergeometrical equation
\begin{equation}
y(1-y)f^{\prime 
\prime}(y)+[(1+2\mu)-y(1+2\mu-2\lambda)]f^{\prime}(y)-[(\mu-\nu-\lambda)
(\mu+\nu-\lambda)]f(y)=0
\end{equation}
where the primes denote derivatives with respect to y and the following 
abbreviations have been used
\begin{eqnarray}
\mu=\frac{ip}{a} \quad , \quad \nu=\frac{ik}{a} \quad , \quad 
\lambda=\frac{iW}{a}\nonumber\\
p^2=(E-W)^2-m^2 \quad , \quad k^2=E^2-m^2
\end{eqnarray}
Note that as we are considering scattering states, $|E|>m$ which 
ensures that $k$ is real, and W is real and positive. $p$ is real for 
$m<E<W-m$ (the Klein range) and $E>W+m$ and imaginary for $W-m<E<W+m$. Of 
principal interest are the energies which lie in the Klein range and 
potentially lead to Klein tunnelling - where fermions can tunnel through 
strong potentials without exponential suppression \cite{KTKP}.
Equation (12) has the general solution
\begin{equation}
f(y)=D_1 \, y^{-2\mu} \, 
{_2F_1}(-\mu-\nu-\lambda,-\mu+\nu-\lambda,1-2\mu;y)+D_2 \, 
{_2F_1}(\mu-\nu-\lambda,\mu+\nu-\lambda,1+2\mu;y)
\end{equation}
So
\begin{eqnarray}
\phi_L(y)=D_1 \, y^{-\mu}(1-y)^{-\lambda} \, 
{_2F_1}(-\mu-\nu-\lambda,-\mu+\nu-\lambda,1-2\mu;y)\nonumber\\
+D_2 \,y^{\mu}(1-y)^{-\lambda}  
{_2F_1}(\mu-\nu-\lambda,\mu+\nu-\lambda,1+2\mu;y)
\end{eqnarray}
For this to be a physically acceptable solution to the problem, it must 
satisfy the appropriate boundary conditions as $x \to -\infty$. The 
solutions as $x \rightarrow -\infty \Rightarrow y \rightarrow - \infty$ 
can be determined using the following formula for the asymptotic 
behaviour of the hypergeometric function \cite{AS}:
\begin{equation}
{_2F_1}(a,b,c;y)=\frac{\Gamma(c)\Gamma(b-a)}{\Gamma(b)\Gamma(c-a)}(-y)^{-a}+
\frac{\Gamma(c)\Gamma(a-b)}{\Gamma(a)\Gamma(c-b)}(-y)^{-b}
\end{equation}
and noting that in the limit $x \to -\infty$, $(-y)^{\mp \nu} 
\rightarrow e^{\pm ik(x+L)}$. Therefore in this limit, $\phi_L(x)$ can be 
written
\begin{equation}
\lim_{x \to -\infty} \, \phi_L(x)=Ae^{ik(x+L)}+Be^{-ik(x+L)}
\end{equation}
From equation (6) the other component, $\chi(x)$ is
\begin{equation}
\chi(x)=\frac{1}{im}[i(E-V(x))\phi(x)-\phi^{\prime}(x)]
\end{equation}
Substituting equation (17) into the above gives us 
\begin{equation}
\lim_{x \to -\infty} \, \chi_L(x)=A \left( \frac{E-k}{m} \right) 
e^{ik(x+L)}+B\left( \frac{E+k}{m} \right) e^{-ik(x+L)}
\end{equation}
where in both cases
\begin{equation}
A=D_1 \, \frac{\Gamma(1-2\mu) \Gamma(-2\nu)}{\Gamma(-\mu-\nu-\lambda) 
\Gamma(1-\mu-\nu+\lambda)}e^{-i \pi \mu}+D_2 \, \frac{\Gamma(1+2\mu) 
\Gamma(-2\nu)}{\Gamma(\mu-\nu-\lambda) \Gamma(1+\mu-\nu+\lambda)}e^{i \pi 
\mu}
\end{equation}
and
\begin{equation}
B=D_1 \, \frac{\Gamma(1-2\mu) \Gamma(2\nu)}{\Gamma(-\mu+\nu-\lambda) 
\Gamma(1-\mu+\nu+\lambda)}e^{-i \pi \mu}+D_2 \, \frac{\Gamma(1+2\mu) 
\Gamma(2\nu)}{\Gamma(\mu+\nu-\lambda) \Gamma(1+\mu+\nu+\lambda)}e^{i \pi 
\mu}
\end{equation}
The choice of combinations of the wave function components (5) can be 
re-written :
\begin{equation}
u_1(x)=\frac{1}{2}(\phi(x)+\chi(x)) \qquad , \qquad 
u_2(x)=\frac{1}{2i}(\phi(x)-\chi(x)) 
\end{equation}
Upon substitution of equations (17) and (19) into the above it can be 
seen that the wave function, $\psi(x)$, comprises of an incident and 
reflected wave far to the left of the barrier which is the desired form to 
establish reflection and transmission amplitudes.

Now consider the solutions for $x>0$. The analysis will differ slightly 
from \cite{DJM} by making a more appropriate substitution to lead to 
the desired transmitted wave function far to the right of the barrier. 
This substitution will also lead to the correct wave functions when the 
bound state solutions are considered. On choosing $z^{-1}=1+e^{a(x-L)}$, 
equation (8) becomes
\begin{equation}
a^2 z(1-z) \frac{d}{dz}\left[ z(1-z) \frac{d \phi_R}{dz} \right]+ [( E 
- Wz)^2-m^2-iaz(1-z)W] \phi_R=0
\end{equation}
Putting $\phi_R=z^{-\nu}(1-z)^{-\mu}g(z)$ and substituting into the 
above gives the hypergeometrical equation
\begin{equation}
z(1-z)g^{\prime 
\prime}(z)+[(1-2\nu)-z(2-2\mu-2\nu)]g^{\prime}(z)-[(1-\mu-\nu-\lambda)
(-\mu-\nu+\lambda)]g(z)=0
\end{equation}
The general solution to the above is
\begin{equation}
g(z)=d_1 \, {_2F_1}(1-\mu-\nu-\lambda,-\mu-\nu+\lambda,1-2\nu;z)+d_2 
\,z^{2\nu}\, {_2F_1}(1-\mu+\nu-\lambda,-\mu+\nu+\lambda,1+2\nu;z)
\end{equation}
So
\begin{eqnarray}
\phi_R=d_1 \,z^{-\nu}(1-z)^{-\mu}\, 
{_2F_1}(1-\mu-\nu-\lambda,-\mu-\nu+\lambda,1-2\nu;z)+\nonumber\\
d_2 \,z^{\nu}(1-z)^{-\mu}\, 
{_2F_1}(1-\mu+\nu-\lambda,-\mu+\nu+\lambda,1+2\nu;z)
\end{eqnarray}
Also as $x \rightarrow \infty$, $z \rightarrow 0$ and $z^{-\nu} 
\rightarrow e^{ik(x-L)}$. Therefore in order to have a plane wave travelling 
to the right as $x \rightarrow \infty$, $d_2=0$. So
\begin{equation}
\phi_R=d_1 \,z^{-\nu}(1-z)^{-\mu}\, 
{_2F_1}(1-\mu-\nu-\lambda,-\mu-\nu+\lambda,1-2\nu;z)
\end{equation}
and
\begin{equation}
\lim_{x \to \infty}\phi_R=d_1 \, e^{-ikL}e^{ikx}
\end{equation}
whilst the other component 
\begin{equation}
\lim_{x \to \infty}\chi_R=d_1 \left( \frac{E-k}{m} \right) \, 
e^{-ikL}e^{ikx}
\end{equation}
In order to find the energy eigenvalues, the wave functions $\phi_L$ 
and $\phi_R$ must be matched at $x=0$. As $x \rightarrow 0$, $y 
\rightarrow 0$ and $z \rightarrow 1$ ($aL \gg 1$), so
\begin{equation}
\phi_L \rightarrow D_1 \, e^{-i \pi \mu}e^{aL \mu}e^{a \mu x}+D_2 \, 
e^{i \pi \mu}e^{-aL \mu}e^{-a \mu x}
\end{equation}
and 
\begin{equation}
\phi_R \rightarrow d_1 \, e^{-a \mu x}e^{aL \mu} 
\frac{\Gamma(2\mu)\Gamma(1-2\nu)}{\Gamma(1+\mu-\nu-\lambda) 
\Gamma(\mu-\nu+\lambda)}+d_1 \, 
e^{a \mu x}e^{-aL \mu} 
\frac{\Gamma(-2\mu)\Gamma(1-2\nu)}{\Gamma(1-\mu-\nu-\lambda) 
\Gamma(-\mu-\nu+\lambda)}
\end{equation}
where use has been made of the following continuation identity for the 
hypergeometric function in $\phi_R$:
\begin{eqnarray}
{_2F_1}(a,b,c;z)=\frac{\Gamma(c)\Gamma(c-a-b)}{\Gamma(c-a) 
\Gamma(c-b)}\,{_2F_1}(a,b,a+b-c+1;1-z)+ \qquad \qquad \qquad \nonumber\\
\qquad \qquad \qquad 
(1-z)^{c-a-b}\frac{\Gamma(c)\Gamma(a+b-c)}{\Gamma(a) \Gamma(b)}\,
{_2F_1}(c-a,c-b,c-a-b+1;1-z)
\end{eqnarray}
Comparing the coefficients of $e^{\pm a \mu x}$ and eliminating $d_1$ 
gives
\begin{equation}
\frac{D_2}{D_1}=e^{-2i\pi \mu}e^{4aL \mu} 
\frac{\Gamma(2\mu)\Gamma(1-\mu-\nu-\lambda)\Gamma(-\mu-\nu+\lambda)}
{\Gamma(-2\mu)\Gamma(1+\mu-\nu-\lambda)\Gamma(\mu-\nu+\lambda)}
\end{equation}
The electrical current density for the  one-dimensional equation (2) is 
defined as 
\begin{eqnarray}
j=\overline{\psi}(x) \gamma_x \psi(x)&=&-\psi(x)^{\dagger} \sigma_2 
\psi(x)\nonumber\\
&=&i\left( u_1^{*}u_2-u_2^{*}u_1 \right) \\
&=&\frac{1}{2}\left( |\phi(x)|^2-|\chi(x)|^2 \right)\nonumber 
\end{eqnarray}
where use has been made of the choice of combinations for the wave 
functions (5). The current as $x \to -\infty$ is $j_L=j_{in}-j_{refl}$ 
where $j_{in}$ is the incident current and $j_{refl}$ is the reflected 
current. Similarly as $x \to \infty$ we have that the current 
$j_R=j_{trans}$ where $j_{trans}$ is the transmitted current. 
Substituting equations (17), (19), (28) and (29) into (34) we find that 
\begin{equation}
\begin{array}{cl}
j_L =&|A|^2\frac{\displaystyle k}{\displaystyle m^2}(E-k) - 
|B|^2\frac{\displaystyle k}{\displaystyle m^2}(E+k)\\[0.3cm]
j_R =&|d_1|^2\frac{\displaystyle k}{\displaystyle m^2}(E-k) \end{array}
\end{equation}
From the conservation of charge we have that $j_L=j_R$ which together 
with the relection coefficient, $R$, and the transmission coefficient, 
$T$
\begin{equation}
R=\frac{j_{refl}}{j_{in}}=\frac{|B|^2}{|A|^2}\left(\frac{E+k}{E-k}\right) 
\qquad , \qquad T= \frac{j_{trans}}{j_{in}} = \frac{|d_1|^2}{|A|^2}
\end{equation}
we obtain the unitarity condition:
\begin{equation}
R+T=1 
\end{equation}
For the purpose of comparison with \cite{DJM} the following expressions 
for the reflection and transmission amplitudes are useful :
\begin{equation}
r= \left( \frac{m+E+k}{m+E-k} \right) e^{-2ikL}\frac{B}{A}\qquad , 
\qquad t=e^{-2ikL}\frac{d_1}{A} 
\end{equation}
The reflection amplitude is found to be
\begin{equation}
r\,=\,\frac{e^{-2ikL}}{\Omega}\frac{(m+E+k)B(2\nu,-\mu-\nu-\lambda)}
{(m+E-k)B(-2\nu,1-\mu+\nu+\lambda)}\times \left[
1-e^{4ipL}\frac{B(2\mu,-\mu-\nu+\lambda) B(2\mu,-\mu+\nu+\lambda)}
{B(-2\mu,\mu+\nu-\lambda) B(-2\mu,\mu-\nu-\lambda)}\right]
\end{equation}
where
\begin{equation}
\Omega=1-e^{4aL\mu} \left[ 
\frac{(\mu+\nu)^2-\lambda^2}{(\mu-\nu)^2-\lambda^2} 
\right]\frac{B^2(2\mu,-\mu-\nu+\lambda)}{B^2(-2\mu,\mu-\nu-\lambda)}
\end{equation}
and $B(a,b)=\frac{\displaystyle \Gamma(a)\Gamma(b)}{\displaystyle 
\Gamma(a+b)}$ is the Beta function.
Note that this equation yields identical results to the $r$ given in 
\cite{DJM}.
The transmission amplitude is found to be
\begin{equation}
t\,=\,\frac{e^{-2ikL+2a \mu 
L}}{\Omega}\left(\frac{(\mu+\nu)^2-\lambda^2}{4 \mu \nu} \right) 
\frac{\Gamma^2(-\mu-\nu-\lambda)\Gamma^2(-\mu-\nu+\lambda)}{\Gamma^2(-2\mu)
\Gamma^2(-2\nu)}
\end{equation}
Using equations (39) and (41), the unitary condition (37) can be 
established; the algebra is quite involved and some care must be taken for 
the two cases where $p$ is real and imaginary.
\begin{figure}[tbph]
\par
\begin{center}
\leavevmode
\includegraphics[width=1\linewidth]{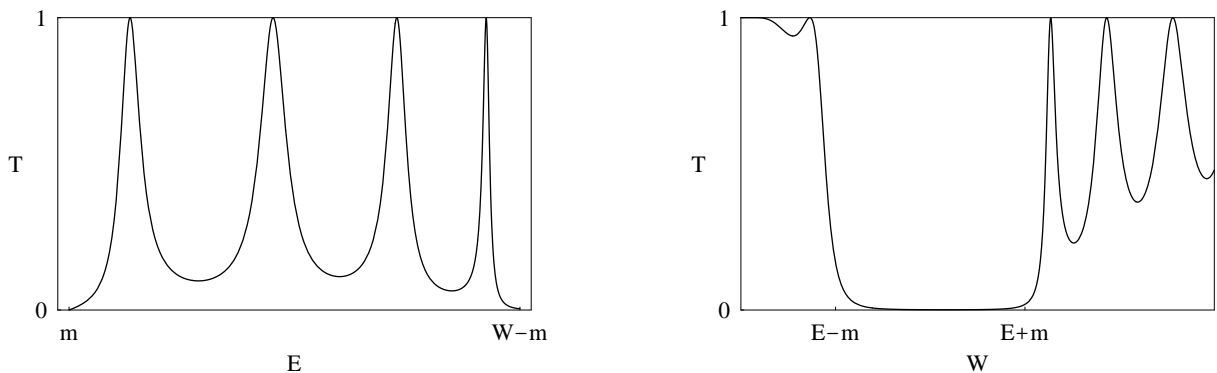} \medskip
\caption{The transmission coefficient for the relativistic Woods-Saxon 
potential barrier. The left hand plot illustrates $T$ for varying 
energy, $E$, with $L=10$, $a=5$, $m=0.4$ and $W=1.2$. The right hand plot is 
for varying barrier height, $W$, with $L=10$, $a=5$, $m=0.2$ and 
$E=2m$. Both plots illustrate the tunnelling without reflection predicted by 
Dosch, Jensen and M\"uller.} 
\end{center}
\end{figure}

In order to establish the condition for a transmission resonance, 
$T=1$, to occur, it is more convenient to look at the reflection amplitude 
for $r=0$. In this instance, only the square-bracketed term on the right 
of (41) needs to be considered; although poles of the gamma-functions 
could make $r$ zero, $\mu$, $\nu$ and $\lambda$ are all purely imaginary 
for energies in the Klein range and do not give the negative integer 
required for this to happen. Consequently, transmission resonances occur 
when
\begin{equation}
1-e^{4ipL}\frac{B(2\mu,-\mu-\nu+\lambda) B(2\mu,-\mu+\nu+\lambda)}
{B(-2\mu,\mu+\nu-\lambda) B(-2\mu,\mu-\nu-\lambda)}=0
\end{equation}
\section*{Bound States}
In order to study the bound states, use can be made of the wave 
function derived for $x>0$, but the analysis can be simplified for $x<0$ by 
making the substitution $y^{-1}=1+e^{-a(x+L)}$ which leads to the 
following equation:
\begin{equation}
a^2 y(1-y) \frac{d}{dy}\left[ y(1-y) \frac{d \phi_L}{dy} \right]+ [( E 
+ Wy)^2-m^2-iay(1-y)W] \phi_L=0
\end{equation}
where $W \rightarrow -W$ in equation (10) for potential wells. Putting 
$\phi_L=y^{\sigma}(1-y)^{\gamma}h(y)$ leads to the hypergeometric 
equation
\begin{equation}
y(1-y)h^{\prime 
\prime}(y)+[(1+2\sigma)-y(2+2\sigma+2\gamma)]h^{\prime}(y)-
[(1+\sigma+\gamma-\lambda)(\sigma+\gamma+\lambda)]h(y)=0 
\end{equation}
where
\begin{eqnarray}
\sigma = \frac{\kappa}{a} \quad , \quad \kappa=m^2-E^2 \quad ,\quad 
\gamma=\frac{ip^{\prime}}{a}
\quad , \quad p^{\prime \, 2}=(E+W)^2-m^2
\end{eqnarray}
As $x \rightarrow -\infty$, $y \rightarrow 0$ so choose the solution
\begin{equation}
h(y)={_2F_1}(1+\sigma+\gamma-\lambda,\sigma+\gamma+\lambda,1+2\sigma;y)
\end{equation}
and therefore
\begin{equation}
\phi_L= 
A^{\prime}y^{\sigma}(1-y)^{\gamma}{_2F_1}(1+\sigma+\gamma-\lambda,\sigma+
\gamma+\lambda,1+2\sigma;y)
\end{equation}
For $x>0$, use equation (26) with $W \rightarrow -W$ so $\lambda 
\rightarrow -\lambda$ and $\mu \to \gamma$ and also $k \rightarrow i \kappa$ 
so $\nu \rightarrow -\sigma$ then choose the solution
\begin{equation}
\phi_R= 
B^{\prime}z^{\sigma}(1-z)^{-\gamma}{_2F_1}(1+\sigma-\gamma+\lambda,\sigma-
\gamma-\lambda,1+2\sigma;z)
\end{equation}
Once again, in order to find the energy eigenvlaues these two wave 
functions must be matched at $x=0$ where both $y$, $z \rightarrow 1$ ($aL 
\gg 1$). By making use of the continuation formula (32) the component 
$\phi$ can be written
\begin{equation}
\phi_L \rightarrow A^{\prime} e^{-\gamma a x}e^{-\gamma a L} 
\frac{\Gamma(-2\gamma) \Gamma(1+2\sigma)}{\Gamma (1-\gamma +\sigma -\lambda) 
\Gamma (-\gamma + \sigma + \lambda)}+
A^{\prime} e^{\gamma a x}e^{\gamma a L} \frac{\Gamma(2\gamma) 
\Gamma(1+2\sigma)}{\Gamma (1+\gamma +\sigma -\lambda) \Gamma (\gamma + 
\sigma + \lambda)}
\end{equation}
and
\begin{equation}
\phi_R \rightarrow B^{\prime} e^{-\gamma a x}e^{\gamma a L} 
\frac{\Gamma(2\gamma) \Gamma(1+2\sigma)}{\Gamma (1+\gamma +\sigma +\lambda) 
\Gamma (\gamma + \sigma - \lambda)}+
B^{\prime} e^{\gamma a x}e^{-\gamma a L} \frac{\Gamma(-2\gamma) 
\Gamma(1+2\sigma)}{\Gamma (1-\gamma +\sigma -\lambda) \Gamma (-\gamma + \sigma 
- \lambda)}
\end{equation}
Comparing terms in $e^{\pm \gamma ax}$ and eliminating $A^{\prime}$ and 
$B^{\prime}$ ultimately leads to
\begin{equation}
\frac{B(-2\gamma,\gamma+\sigma-\lambda)^2}{B(2\gamma,-\gamma+\sigma+\lambda)^2}
=e^{4\gamma 
a L} \frac{(\sigma-\gamma)^2-\lambda^2}{(\sigma+\gamma)^2-\lambda^2}
\end{equation}
So
\begin{equation}
\frac{B(-2\gamma,\gamma+\sigma-\lambda)}{B(2\gamma,-\gamma+\sigma+\lambda)}
=\pm 
e^{2\gamma a L} 
\sqrt{\frac{(\sigma-\gamma)^2-\lambda^2}{(\sigma+\gamma)^2-\lambda^2}}
\end{equation}
where the even solutions are determined by the positive square root and 
the odd solutions by the negative square root. These equations need to 
be solved numerically to find the energy eigenvalues for the bound 
states. It is also possible to plot the upper and lower components of the 
bound state wave function $\psi(x)$:
\begin{figure}[tbph]
\par
\begin{center}
\leavevmode
\includegraphics[width=0.6\linewidth]{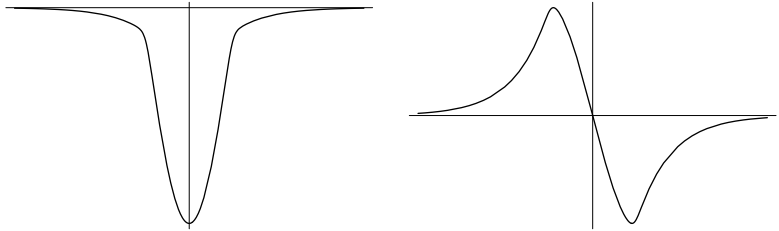} \medskip
\caption{The upper (left) and lower (right) components of the first 
even bound state wave function for the sub-critical Woods-Saxon potential 
well where $a=10$, $L=2$, $m=1$ and $W=2$. The energy of the bound 
state is $E=0.759$.}
\end{center}
\end{figure}

\section*{Supercriticality}
As the potential well deepens for increasing $W$, the energy eigenvalue 
of any given bound state will also decrease. When this energy reaches 
$E = -m$, the bound state merges with the negative energy continuum and 
the potential is said to be supercritical. When $E \to -m$, $\kappa \to 
0$ and consequently $\sigma \to 0$. Writing
\begin{equation}
\gamma_c=\frac{i}{a}p_c \qquad , \qquad p_c^2=W^2-2mW
\end{equation}
the energy eigenvalue equation (51) becomes
\begin{equation}
\frac{B(-2\gamma_c,\gamma_c-\lambda)^2}{B(2\gamma_c,-\gamma_c+\lambda)^2}
=e^{4\gamma_c 
a L}
\end{equation}
Once again the even and odd supercritical energy eigenvalues can be 
determined by taking the positive and negative square roots respectively.

\section*{Square Well Limit : Bound States}
Using
\begin{equation}
B(a,b)=\left( \frac{a+b}{b} \right)
\left( \frac{a+b+1}{a} \right) B(a+1,b+1)
\end{equation}
we find that
\begin{equation}
B(\mp 2 \gamma,\pm \gamma +\sigma \mp \lambda)= \mp \frac{(\mp 
\gamma+\sigma \mp \lambda)(1 \mp \gamma+\sigma \mp \lambda)}{2\gamma(\pm 
\gamma+\sigma \mp \lambda)}B(1 \mp 2\gamma,1 \pm \gamma +\sigma \mp \lambda)
\end{equation}
Therefore in the square well limit $a \to \infty$, 
$\gamma,\sigma,\lambda \to 0$ and
\begin{equation}
\frac{B(-2 \gamma,\gamma +\sigma -\lambda)}{B(2 \gamma,-\gamma +\sigma 
+\lambda)} \to
-\frac{(-\gamma+\sigma-\lambda)(-\gamma+\sigma+\lambda)}{(\gamma+\sigma-\lambda)
(\gamma+\sigma+\lambda)}
\end{equation}
where we have used $B(1,1)=1$. Substituting this into (51) we obtain
\begin{equation}
e^{4ip^{\prime}L} = 
\frac{(\sigma-\gamma)^2-\lambda^2}{(\sigma+\gamma)^2-\lambda^2}
= \frac{(\kappa-ip^{\prime})^2+W^2}{(\kappa+ip^{\prime})^2+W^2}
\end{equation}
Rationalising we find that 
\begin{equation}
e^{4ip^{\prime}L} =\left( \frac{\kappa^2-p^{\prime 
2}+W^2}{2mW}-\frac{i\kappa p^{\prime}}{mW} \right)^2
\end{equation}
Choosing the positive root and solving for the real and imaginary parts 
one eventually obtains
\begin{equation}
tan \, 2p^{\prime} L =\frac{2 \kappa p^{\prime}}{p^{\prime 
2}-\kappa^2-W^2}
\end{equation}
which after much laborious algebra gives for the even solutions
\begin{equation}
tan \, p^{\prime} L = \frac{mW+\kappa^2-EW}{\kappa p^{\prime}}
= \sqrt\frac{(m-E)(E+W+m)}{(m+E)(E+W-m)}
\end{equation}
and for the odd solutions
\begin{equation}
tan\,p^{\prime} L = -\frac{mW-\kappa^2+EW}{\kappa p^{\prime}}
= -\sqrt\frac{(m+E)(E+W-m)}{(m-E)(E+W+m)}
\end{equation}
(these are precisely the equations for even and odd bound states in the 
square well \cite{CDI} with $W=V$).

\section*{Zero Momentum Resonances and Supercriticality}

It was first pointed out in \cite{PDKMSc} that the conditions for 
supercriticality and zero-momentum resonances were the same for the square, 
Gaussian and Woods-Saxon potentials. Indeed for the square potential 
well, $V(x)$, where $V(x)=0$ for $|x| \geq a$ and $V(x)=-U$ $\leq 0$ for 
$|x| <a$, the condition for supercriticality is $2p^{\prime}a=N \pi$ 
where $p^{\prime 2} =U^2-2mU$ \cite{KTKP}. The Dirac equation (4) is 
invariant under charge conjugation: that is to say under the transformation
\begin{equation}
E\rightarrow -E\quad U\rightarrow -U\quad u_1\rightarrow u_2\quad 
u_2\rightarrow u_1
\end{equation}
Consequently the condition for a supercritical particle at $E=-m$ in a 
square well is the same as that for a supercritical antiparticle at 
$E=m$ in a square barrier. It can also be shown that for a transmission 
resonance to occur, $2pa=N \pi$ where $p^2=(E-V)^2-m^2=k^2+V^2-2VE$. In 
the zero-momentum limit, $k \to 0$, this is seen to be identical to the 
condition for a supercritical antiparticle. In other words when a 
potential well of finite range is strong enough to contain a supercritical 
state, then a particle of arbitrarily small momentum will be able to 
tunnel right through the potential barrier created by inverting the well 
without reflection \cite{DKC}. From (42), the condition for a 
transmission resonance to occur for the Woods-Saxon potential is 
\begin{equation}
e^{4ipL}=\frac{B(-2\mu,\mu+\nu-\lambda) 
B(-2\mu,\mu-\nu-\lambda)}{B(2\mu,-\mu-\nu+\lambda) B(2\mu,-\mu+\nu+\lambda)}
\end{equation}
When $E \to m$ (for a low-momentum particle), we find that $p^2 \to 
W^2-2mW=p_c$ so that $\mu \to \gamma_c$ and also $k \to 0 \Rightarrow \nu 
\to 0$. So (64) becomes
\begin{equation}
e^{4ip_c L}= 
\frac{B(-2\gamma_c,\gamma_c-\lambda)^2}{B(2\gamma_c,-\gamma_c+\lambda)^2}
\end{equation}
As expected, this is precisely the condition required for the potential 
barrier to be supercritical (54) once the \lq flipping\rq\, procedure 
on the potential well as described above is considered. The components 
of the zero-momentum resonance/half-bound state wave function, 
$\psi(x)$, can be plotted and have the following appearance:
\begin{figure}[tbph]
\par
\begin{center}
\leavevmode
\includegraphics[width=0.6\linewidth]{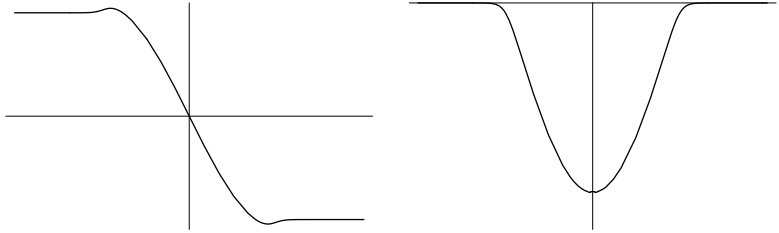} \medskip
\caption{The upper (left) and lower (right) components of the first 
zero momentum wave function for the supercritical Woods-Saxon potential 
barrier where $a=10$, $L=2$, $m=1$ and $W=2.274$.}
\end{center}
\end{figure}
Comparison with the zero momentum wave functions for the square and 
Gaussian barriers \cite{DKC} highlights the existence of two turning 
points which occur in the top component of the wave function for the smooth 
Gaussian potential only. Once again these correspond to the points $\pm 
x_K$ where $V(\pm x_K)=E+m=2m$ at zero momentum and are not manifest 
for the square barrier whose walls are discontinuous. 

\section*{Acknowledgement}
The author wishes to thank Professor Norman Dombey for many helpful 
discussions and suggested improvements to the paper.

\end{document}